\documentclass[sigconf]{acmart}
\renewcommand\footnotetextcopyrightpermission[1]{} 
\def\BibTeX{{\rm B\kern-.05em{\sc i\kern-.025em b}\kern-.08emT\kern-.1667em\lower.7ex\hbox{E}\kern-.125emX}}
    
\copyrightyear{2019} 
\acmYear{2019} 
\setcopyright{acmlicensed}
\acmConference[IH\&MMSec '19]{ACM Information Hiding and Multimedia Security Workshop}{}{}
\acmPrice{15.00}
\acmDOI{10.1145/3335203.3335738}
\acmISBN{978-1-4503-6821-6/19/06}

\usepackage{blindtext, graphicx}
\usepackage{xcolor}
\usepackage[normalem]{ulem}
\usepackage{tabu}
\usepackage{multirow}
\usepackage{hhline}
\usepackage{arydshln}
\usepackage{graphics}
\usepackage{amsmath}
\usepackage{mathrsfs}
\usepackage[caption=false]{subfig}
\usepackage{graphicx}

\def\CA{\mathcal{C}_A}
\def\SA{\mathcal{S}_A}
\def\SB{\mathcal{S}_B}
\def\DB{\mathcal{D}_B}

\def\train{^\mathrm{train}}
\def\test{^\mathrm{test}}

\def\TN{\mathrm{TN}}
\def\TP{\mathrm{TP}}
\def\FN{\mathrm{FN}}
\def\FP{\mathrm{FP}}
\def\INC{\mathrm{INC}}
\def\Err{\mathrm{Err}}
\def\pred{_\mathrm{pred}}

\begin{document}

{\Huge
This is a preprint of the paper:\\

Daniel Lerch-Hostalot, David Meg\'{\i}as, \emph{"Detection of Classifier Inconsistencies in Image Steganalysis"}, Proceedings of the ACM Workshop on Information Hiding and Multimedia Security, July 2019, Pages 222-229. ISBN: 978-1-4503-6821-6. \hyperlink{https://doi.org/10.1145/3335203.3335738}{https://doi.org/10.1145/3335203.3335738}. 
}
\newpage

\title{Detection of Classifier Inconsistencies in Image Steganalysis}



\author{Daniel Lerch-Hostalot \and David Meg\'{\i}as}
\affiliation{%
  \institution{Internet Interdisciplinary Institute (IN3), Universitat Oberta de Catalunya (UOC),\\ CYBERCAT-Center for Cybersecurity Research of Catalonia}
 \streetaddress{Avgda. Carl Friedrich Gauss, 5,}
 \city{Castelldefels}
 \state{Barcelona, Spain}
 \postcode{08860}
}
\email{{dlerch,dmegias}@uoc.edu}




\begin{abstract}
In this paper, a methodology to detect inconsistencies in {clas\-si\-fi\-ca\-tion-based} image steganalysis is presented. The proposed approach uses two classifiers: the usual one, trained with a set formed by \emph{cover} and \emph{stego} images, and a second classifier trained with the set obtained after embedding additional random messages into the original training set. When the decisions of these two classifiers are not consistent, we know that the prediction is not reliable. The number of inconsistencies in the predictions of a testing set may indicate that the classifier is not performing correctly in the testing scenario. This occurs, for example, in case of cover source mismatch, or when we are trying to detect a steganographic method that the classifier is no capable of modelling accurately. We also show how the number of inconsistencies can be used to predict the reliability of the classifier (classification errors).
\end{abstract}



\maketitle

\section{Introduction}

Steganography is a collection of techniques to embed secret data into apparently innocent objects. Nowadays, these objects are {main\-ly} digital media, and  the most common carriers for steganography are digital images because of their widespread use. On the other hand, steganalysis refers to different techniques used to detect messages previously hidden using steganography. 

Most steganalytic methods in the state of the art use \emph{machine learning} \cite{Fridrich:2012:RM, Denemark:2014:selchannel, Boroumand:2019:SRNet}. In machine learning-based steganalysis, firstly, a (training) set of known cover and stego images is used to train a classifier. Later on, this classifier is used to predict the images of a testing set as cover or stego. 

This approach works very well in laboratory conditions, that is, if the set of training images is similar to that of the testing images used by the steganographer to hide secret data. However, in the real world, the set of media used by the steganographer might be quite different from that used to train the classifier \cite{Ker:2013:real_world}. This occurs, for example, when the images in the testing set are not well represented in the training set. Some examples of this mismatch occur if the testing images are taken using a different camera or resolution; if they are compressed, zoomed or improved through filters; or if they were taken in very different conditions. In steganalysis, this problem is known as \emph{cover source mismatch} (CSM) and was initially reported in \cite{Cancelli:2008:csm}.
There are other situations that lead to inaccurate predictions. For example, \emph{stego source mismatch} (SSM) occurs when some embedding parameter, such as the exact payload \cite{Pevny:2011}, differ between the training and the testing datasets.

Different approaches to the CSM problem have been proposed. During the BOSS competition \cite{Bas:2011:boss}, some participants tried an approach called ``training on a contaminated database'', which consists in denoising images from the testing set and including them in the training set \cite{Kurugollu:2011:hugo}. A different approach is to make the training set as complete as possible. In \cite{Lubenko:2012}, the authors trained a classifier with a huge variety of images. Due to the high time and memory requirements, this was carried out using on-line classifiers. In \cite{Fridrich:2014:csm_mitig}, three different strategies are presented: (1) training with a mixture of cover sources; (2) using different classifiers trained with different sources and testing with the closest source; and (3) taking the second approach but testing each image separately using the closest source. The \emph{islet approach} \cite{Pasquet:2014}, introduces a pre-processing step consisting in organizing the images in clusters and assigning a steganalyzer to each cluster. In \cite{Xu:2015:csm},  a scheme to efficiently construct a large and representative training set is proposed. 
Finally, in \cite{Lerch-Hostalot:2015}, an unsupervised steganalytic method was proposed that does not require a training set, bypassing the CSM problem.

In supervised machine learning-based steganalysis, we need a database of images to construct the training set and a validation set to determine the classification accuracy results. The creation of this database is a fundamental part of the process. Usually, this is carried out by collecting pictures taken with different cameras and models, taken in different lighting conditions, compressed with different algorithms and compressing ratios, processed with different filters, modified by optical or digital zoom, etc. If we create a database with pictures taken by a team of people with their cameras and with a specific set of filters, zooms, compression algorithms and so on, this will be a biased procedure, and such a selection can never represent the entire population of possible images. Even if we decide to download random images from the Internet, the combination of cameras, models, filters, compression ratios, light conditions, and so on, is too large and it is almost impossible to obtain a representative enough dataset. 

The current approach to solve this problem consist in building a very large and heterogeneous dataset, as described in \cite{Cogranne:2018:alaska}. By taking such an approach, the steganalyst expects that the trained classifier (usually a convolutional neural network) will learn a collection of features that are universal to all images and, consequently, good enough to classify images taken under conditions different from those of the training set. Although this approach is often successful, we think that it is convenient to test other solutions to the problem. In this paper, we explore an alternative approach.

The proposed approach is based on the ideas of \cite{Lerch-Hostalot:2015}, which are extended here to detect samples that lead to classification inconsistencies. Thus, the suggested method stems from obtaining additional training and testing sets by sequential random data embedding. Those additional sets are used here to detect inconsistencies in the classification. We present a method in the context of batch steganography \cite{Ker:2006:batch}, that is, when we are analyzing a set of images from a suspicious source.

The rest of this paper is organized as follows. Section \ref{sec:preliminaries} introduces some relevant concepts that are used in the proposed method. Section \ref{sec:proposed} presents the proposed method. Experimental results obtained with the proposed method for different image databases, embedding algorithms and steganalytic classifiers (including CSM cases) are presented in Section \ref{sec:experimental}. Finally, Section  \ref{sec:conclusion} summarizes the conclusions and suggests some directions for further research.

\section{Preliminaries}
\label{sec:preliminaries}

We consider a targeted scenario in which the embedding algorithm and the approximate embedding rate --but not the secret key-- are assumed to be known (at least approximately) by the steganalyst. Using the same steganographic algorithm and embedding bit rate, new (random) data can be hidden into any image with a different (random) secret key. Given a set of cover images, they can be used to build a training database by embedding random data 
to obtain a training database consisting of a half of cover and a half of stego images. This set is called $A\train$.  If a feature extraction-based machine learning algorithm is applied (not all the machine learning algorithms need feature extraction \cite{Boroumand:2019:SRNet}), we need to extract the features of the images. The usual methodology in machine learning-based steganalysis is to use the set $A\train$ to train a classifier. Then, this classifier can be used to classify a testing set $A\test$, that is, a set of images for which we do not have \textit{a priori} information whether they are cover or stego. 

The proposed methodology uses an additional set, $B\train$, defined as suggested in \cite{Lerch-Hostalot:2015}. The set $B\train$ is the result of hiding random data into all the images of the set $A\train$ using the targeted embedding algorithm, the approximate embedding bit rate and random keys. As a result, we have a set $A\train$, which contains cover and stego images, and a set $B\train$, which contains stego and ``double stego'' images. 

Now, we introduce the following notation: let $\alpha_i$ be a sample from the set $A\train$ and $\beta_i$ be the corresponding sample from the set $B\train$, whereby $\beta_i=\mathrm{Embed}(\alpha_i,\mathrm{Bitrate})$. ``Embed'' stands for embedding a random message, using a random key and the targeted steganographic algorithm, and ``Bitrate'' is the known (or approximated) embedding bit rate.

Similarly, from the images in $A\test$, we build an additional set $B\test$ following the same procedure:  $a_i$ and $b_i$ stand for samples of the training sets $A\test$ and $B\test$, respectively, with \linebreak $b_i=\mathrm{Embed}(a_i,\mathrm{Bitrate})$. The only difference with respect to the training sets is that we do not know the classes (labels) of the images of the testing sets. In other words, we know that $A\test$ possibly contains both cover and stego images, and that $B\test$ possibly contains stego and ``double stego'' images, but we do not know the class of each image.

Finally, we assume that there is a classifier ${\hat f}_A$ (and a feature extractor if needed) that can split images into cover ($\CA$) and stego ($\SA$) classes with an acceptable probability of error. Similarly, we assume that we have another classifier ${\hat f}_B$ that can split images into stego ($\SB$) and ``double stego'' ($\DB$) classes,

\begin{table}[ht]
\begin{center}

\caption{Number of pixels modified by $\pm 1$ and $\pm 2$ for 1,000 images after the first and the second embeddings. ``ALGO''/``BR'': embedding algorithm and bit rate (bpp)}


%
%
%

\label{tab:pmn}
\begin{tabular}{cc|rr|rr}
\hline 
\multicolumn{2}{c|}{} &
\multicolumn{2}{c|}{1 embedding} &
\multicolumn{2}{c}{2 embeddings} \\
\textbf{\scriptsize ALGO} & 
\textbf{\scriptsize BR} & 
\multicolumn{1}{c}{\textbf{\scriptsize $\pm1$}} & 
\multicolumn{1}{c|}{\textbf{\scriptsize $\pm2$}} &
\multicolumn{1}{c}{\textbf{\scriptsize $\pm1$}} & 
\multicolumn{1}{c}{\textbf{\scriptsize $\pm2$}}\\
\hline

HILL & 0.4 & 22,582,706 & 0 & 33,993,485 & 2,819,705 \\
HILL & 0.2 & 9,897,485  & 0 & 16,112,459 & 933,376 \\

UNIWARD & 0.4 & 19,509,940 & 0 & 32,748,639 & 1,563,635 \\
UNIWARD & 0.2 & 8,523,446  & 0 & 15,139,023 & 477,200 \\

LSBM & 0.2 & 27,528,954 & 0 & 49,277,814 & 1,439,498 \\
\hline
\end{tabular}
\end{center}
\end{table}

\newcommand{\mycaption}[1]{\stepcounter{figure}\raisebox{-7pt}
  {\footnotesize Fig. \thefigure.\hspace{3pt} #1}}

\begin{figure*}[ht]
\centering

  \subfloat[$A\train$ and $B\train$ sets]{\label{fig:nasd_AB}
  \ifpdf \includegraphics[width=0.35\textwidth]{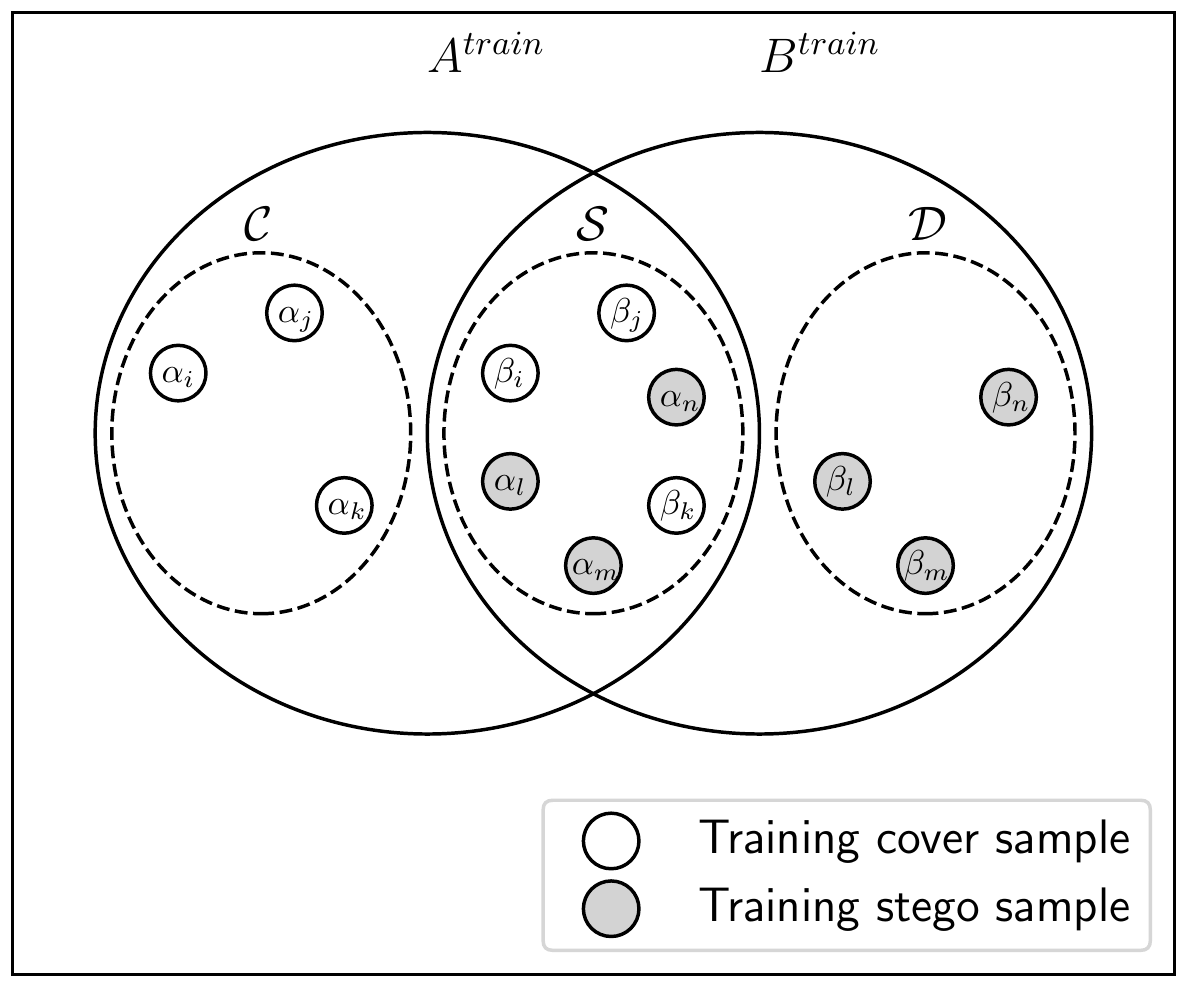}
  \else \includegraphics[width=0.35\textwidth]{nasd_AB.eps}
  \fi}
  \subfloat[Consistent predictions]{\label{fig:nasd_aligned}
  \ifpdf \includegraphics[width=0.35\textwidth]{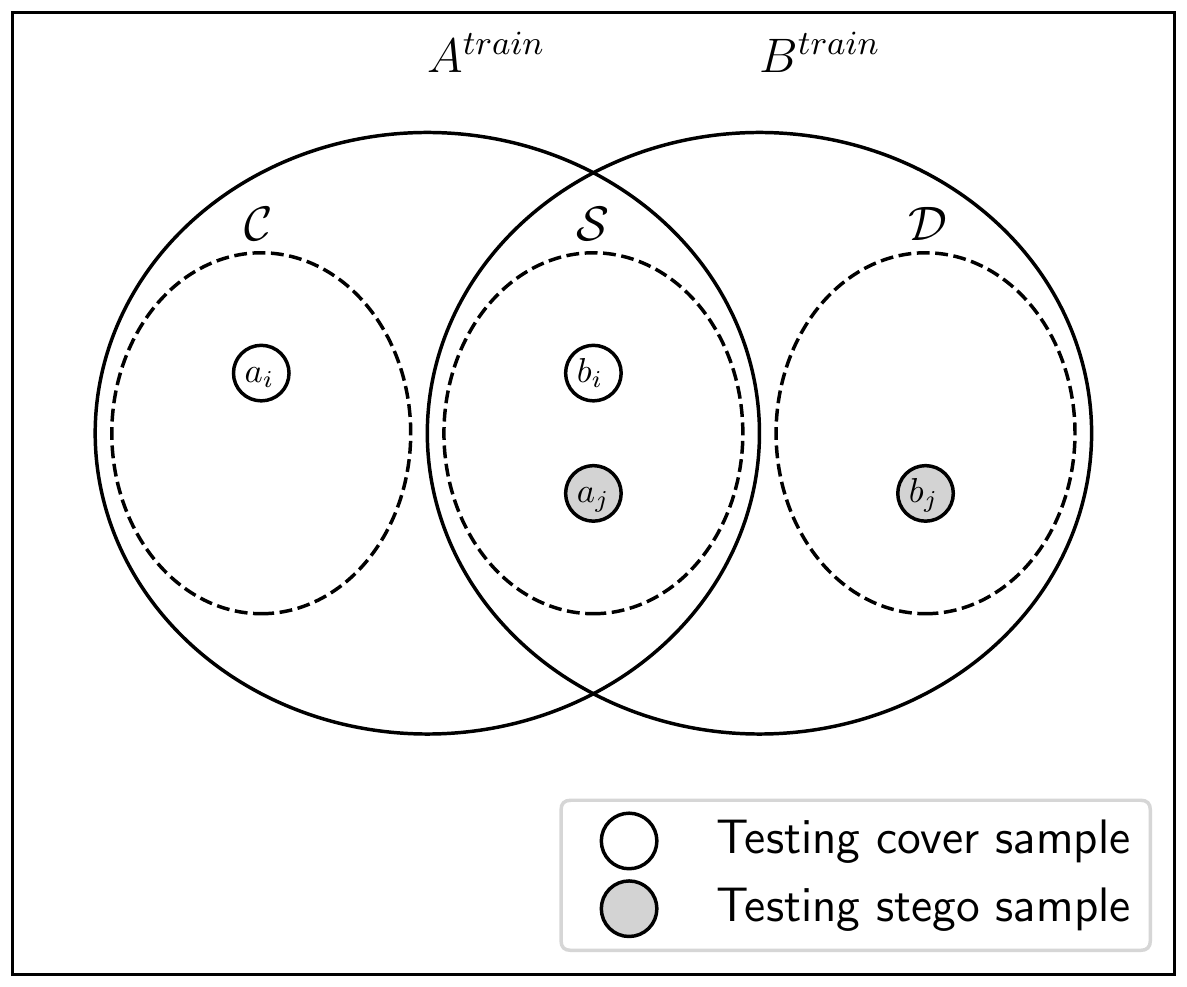}
  \else \includegraphics[width=0.35\textwidth]{nasd_aligned.eps}
  \fi} \\
  
  \subfloat[Undetectable inconsistencies]{\label{fig:nasd_undetect}
  \ifpdf \includegraphics[width=0.31\textwidth]{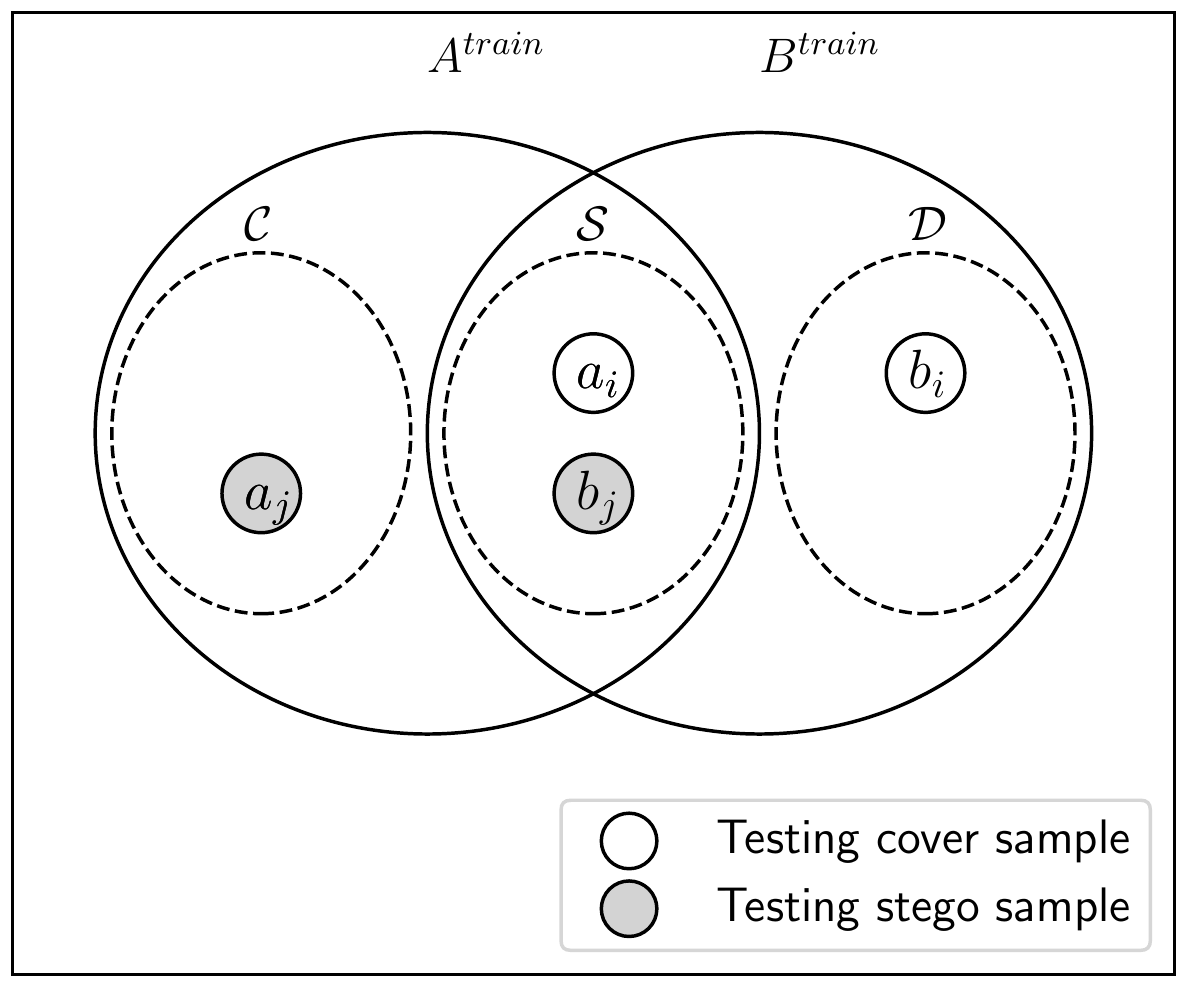}
  \else \includegraphics[width=0.31\textwidth]{nasd_undetect.eps}
  \fi} 
  \subfloat[Some $F_1$-detectable inconsistencies]{\label{fig:nasd_f1detect}
  \ifpdf \includegraphics[width=0.31\textwidth]{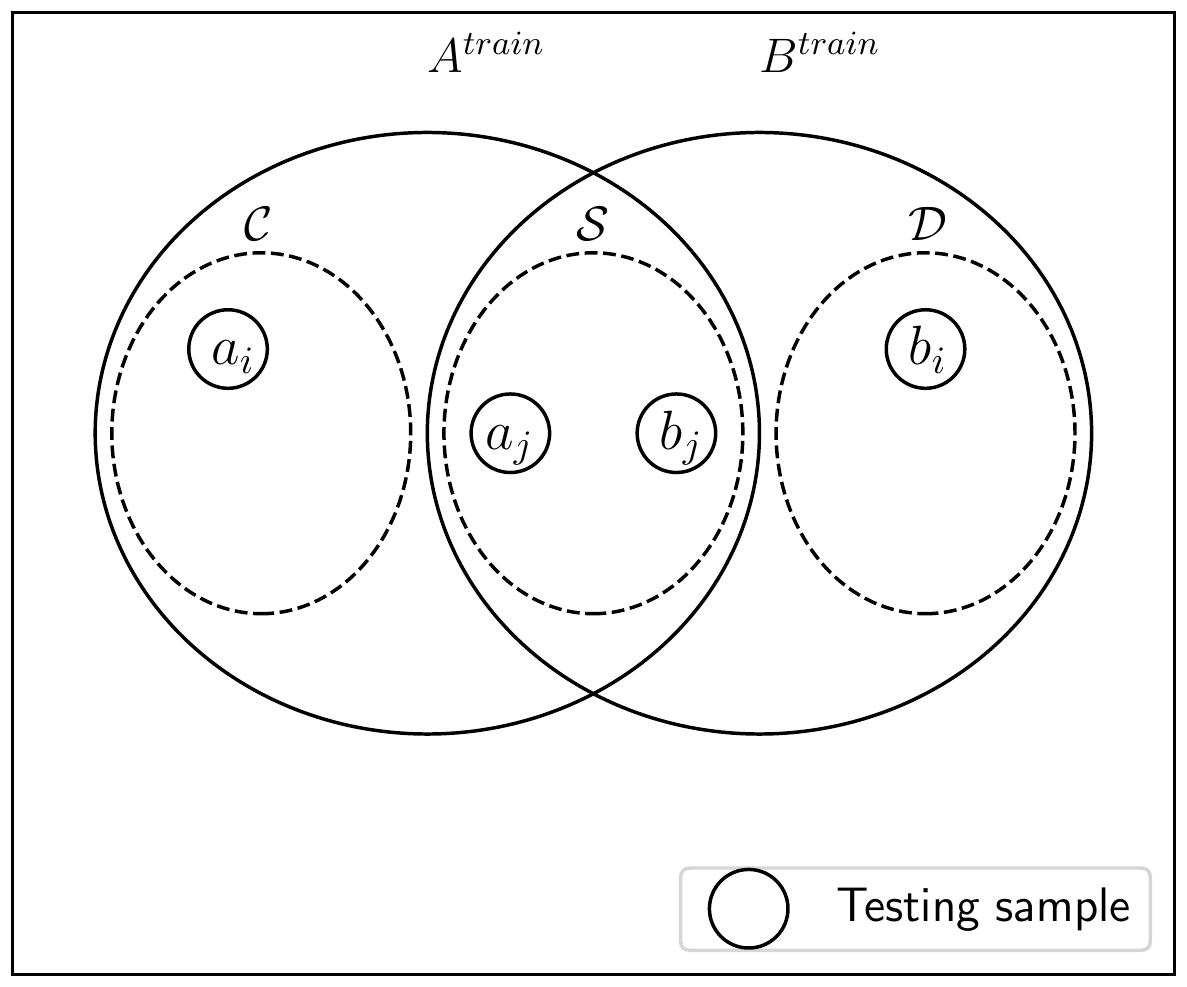}
  \else \includegraphics[width=0.31\textwidth]{nasd_f1detect.eps}
  \fi}
  \subfloat[$F_2$-detectable inconsistencies]{\label{fig:nasd_f2detect}
  \ifpdf \includegraphics[width=0.31\textwidth]{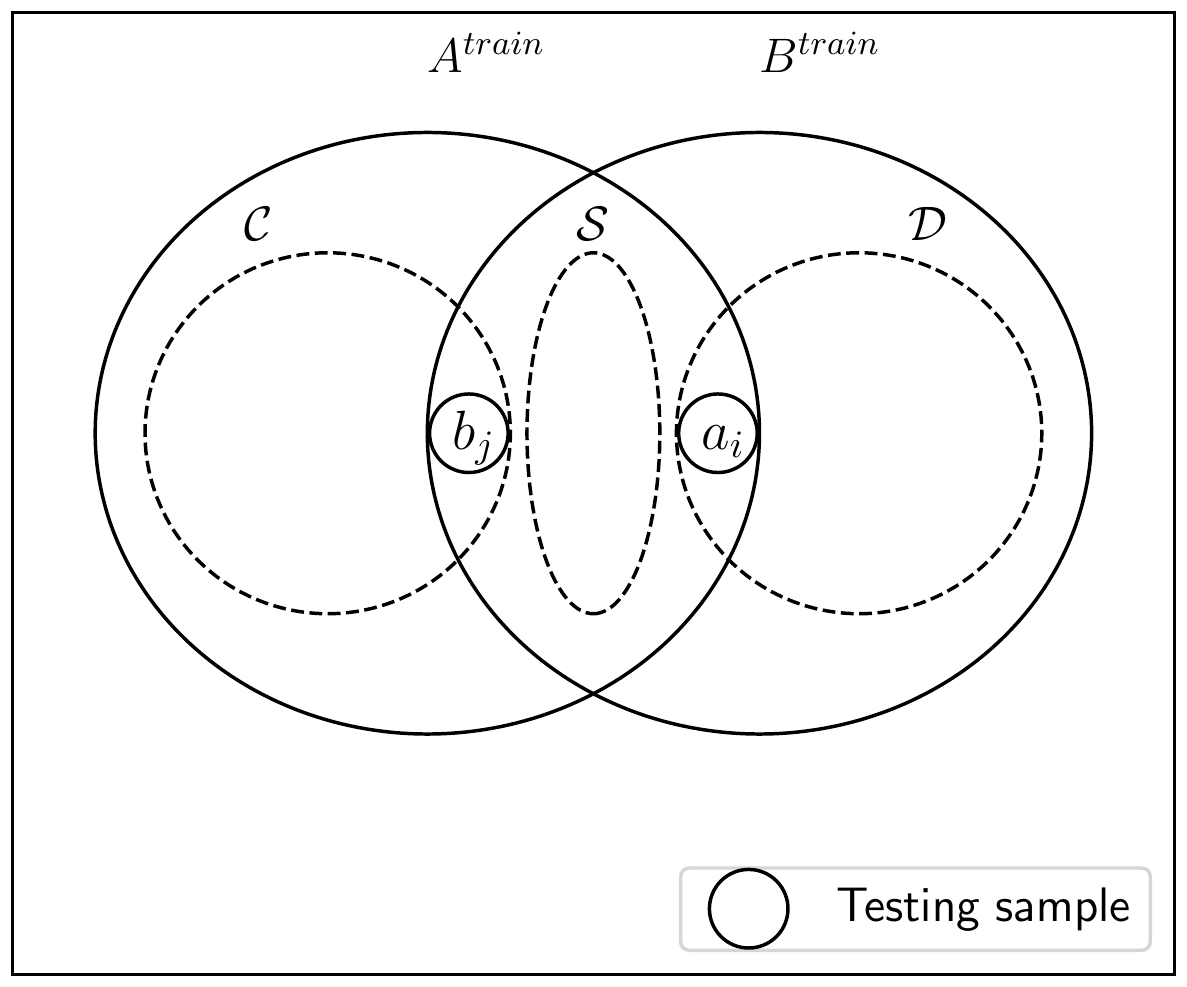}
  \else \includegraphics[width=0.31\textwidth]{nasd_f2detect.eps}
  \fi} 
  \quad\\
  
  \centering\mycaption{Graphical representation of the proposed method}
  \label{fig:ATS_method}
\end{figure*}

The assumption of the existence of $\hat{f}_B$ that can split images into stego ($\SB$) and ``double stego'' ($\DB$) classes needs some discussion. After all, both $\SB$ and $\DB$ classes are formed by stego images with more or less information hidden into them. Usually, in the spatial domain, steganographic algorithms embed information by carrying out a $\pm 1$ operation in some specific pixels. In a second embedding, the steganographic algorithm may choose an already modified pixel to hide additional information. Thus, in addition to $\pm 1$ changes, there is some probability of a $\pm 2$ operation for a few pixels. In the particular case of adaptive embedding, a probability map indicates the areas that will be selected for hiding information. These probability maps are quite similar for the cover and stego versions of the same image. Therefore, the algorithm tends to hide information in the same pixels. This increases the differences between the features of stego and ``double stego'' images. These differences become more detectable with greater embedding ratios. One can conclude that the patterns of pixels (and its neighbours) generated after the second embedding will be different from those of a single embedding, and a good enough classifier can take advantage of those differences. In Table \ref{tab:pmn}, the number of $\pm1$ and $\pm2$ variations after embedding messages with different algorithms and bit rates in 1,000 images randomly selected from the BOSS base are shown. We have included an experiment with LSB matching to show that even with non-adaptive steganography, the number of $\pm2$ variations is not negligible. It can be observed that the number of $\pm 2$ variations is relatively large and and that the final number of $\pm 1$ changes is higher after the second embedding. These two factors help the classifier to split stego and ``double stego'' images. Even in the case of LSB matching, which does not use a probability map that forces the algorithm to hide data in the same positions, the number of $\pm 2$ variations is quite high. To analyze the relevance of $\pm 2$ changes for the classification, we trained a classifier $\hat{f}_B$ using the BOSS database and the HILL embedding algorithm with a bit rate of $0.4$ bits per pixel (bpp). We obtained a classification error of $0.2770$. Next, we trained the classifier with the same images after replacing the $\pm 2$ changes by their respective $\pm 1$ values. In this case, we obtained an error of $0.3160$, which is slightly worse. Therefore, it can be seen that, although the influence of $\pm 2$ changes in the classification is very significant, the increase in the number of $\pm 1$ changes is enough to split both classes.

\section{Proposed method}
\label{sec:proposed}

This section outlines the method proposed to detect the number of inconsistencies occurred during classification. We also propose a mechanism to predict the error incurred by the classifier based on the number of such inconsistencies.

Usually, in machine learning-based steganalysis, when we know the embedding algorithm and the approximate embedding rate, the testing set is expected to be classified with some (hopefully low) probability of error \cite{Fridrich:2012:RM}. If this idea is extended to the sets introduced in the previous section, we have a tool that can be used to detect inconsistencies in classification. Since there is a bijection between the elements of $A\test$ and $B\test$, if an image is classified as cover using ${\hat f}_A$, the corresponding image in $B\test$ should be classified as stego (not ``double stego'') using ${\hat f}_B$. Similarly, if an image in $A\test$ is classified as stego using ${\hat f}_A$, we expect the same image in $B\test$ to be classified as ``double stego'' if we use ${\hat f}_B$. We call ``inconsistency'' to a classification result that does not meet these requirements.

The classification constraints described above are used to define \emph{filters}. The filter described in the previous paragraph is denoted as $F_1$, and consists in classifying $a_i$ using ${\hat f}_A$ and $b_i$ using ${\hat f}_B$ to check whether the two classification results are consistent:
$$
F_1(i)\equiv\left\{
\begin{array}{ll}
\text{If }{\hat f}_A(a_i)=\SA, & \text{If }({\hat f}_B(b_i) \neq \DB)\\ 
& \text{ then output ``inconsistency''}, \\
&\\
\text{Otherwise, } & \text{If }({\hat f}_B(b_i) \neq \SB)\\ 
& \text{ then output ``inconsistency''}.
\end{array}\right.
$$

In Fig.\ref{fig:nasd_AB}, we can see a graphical representation of the classes. A consistent classification is represented in  Fig.\ref{fig:nasd_aligned}. In Fig.\ref{fig:nasd_f1detect}, we can see different inconsistencies that can be detected with the filter $F_1$.

Now, we can consider the case in which ${\hat f}_A$ is used to classify $a_i\in A\test$ and the classification result is stego. If we classify $a_i\in A\test$ using ${\hat f}_B$, we expect that it is classified also as stego (not as ``double stego''). If $a_i$ is classified as stego using ${\hat f}_A$ and as ``double stego'' by ${\hat f}_B$, there is an inconsistency. In fact, if $a_i$ corresponds to a cover image, it would not be consistent that $a_i$ is classified by ${\hat f}_B$ as ``double stego'' either. Hence, for any image $a_i$ in $A\test$, the output of ${\hat f}_B$ taking $a_i$ as input must be always stego and never ``double stego''. The same idea can be applied to inconsistencies for the set $B\test$. If we use ${\hat f}_A$ to classify any sample $b_i\in B\test$, we expect $b_i$ to be classified  as stego (and never as cover). This kind of filter is denoted as $F_2$ (represented in Fig.\ref{fig:nasd_f2detect}):
$$
F_2(i)\equiv\left\{
\begin{array}{ll}
\text{If }{\hat f}_B(a_i) \neq \SB, & \text{ then output ``inconsistency''}, \\
&\\
\text{If }{\hat f}_A(b_i) \neq \SA, & \text{ then output ``inconsistency''}.
\end{array}\right.
$$

\begin{table*}[ht]
\begin{center}

\caption{Classification results for standard RM+EC (without filters) and for the proposed method (with filters). ``N'': number of experiment,  ``ALGO'': name of the embedding algorithm and embedding bit rate (bpp), ``DBs'': training/testing databases, ``C/S'': number of cover and stego images in the testing set , ``CLF'': classification method used, ``Err'': true classification error,  \{``TP'', ``TN'', ``FP'', ``FN''\}: true and false positives and negatives, ``INC'': number of inconsistencies,  ``INC$_\mathbf{\mathrm C}$'': number of inconsistencies for images predicted as cover, and  ``INC$_\mathbf{\mathrm S}$'': number of inconsistencies for images predicted as stego}
\label{tab:experiments}
\resizebox{0.975\textwidth}{!}{\begin{tabular}{lllcc|>{\bfseries}rrrrr|>{\bfseries}rrrrrrrrr}

\hline 

\multicolumn{5}{c|}{} &
\multicolumn{5}{c|}{\textbf{STANDARD}} &
\multicolumn{9}{c}{\textbf{PROPOSED}} \\

\textbf{\scriptsize N} &
\textbf{\scriptsize ALGO} & 
\textbf{\scriptsize DBs} &
\textbf{\scriptsize C/S} &
\textbf{\scriptsize CLF} &
\multicolumn{1}{c}{\textbf{\scriptsize Err}} & 
\multicolumn{1}{c}{\textbf{\scriptsize $\TP$}} & 
\multicolumn{1}{c}{\textbf{\scriptsize $\TN$}} & 
\multicolumn{1}{c}{\textbf{\scriptsize $\FP$}} & 
\multicolumn{1}{c|}{\textbf{\scriptsize $\FN$}} & 
\multicolumn{1}{c}{\textbf{\scriptsize Err$_{pred}$}} &
\multicolumn{1}{c}{\textbf{\scriptsize Err}} &
\multicolumn{1}{c}{\textbf{\scriptsize $\TP$}} & 
\multicolumn{1}{c}{\textbf{\scriptsize $\TN$}} & 
\multicolumn{1}{c}{\textbf{\scriptsize $\FP$}} & 
\multicolumn{1}{c}{\textbf{\scriptsize $\FN$}} &
\multicolumn{1}{c}{\textbf{\scriptsize $\INC$}} &
\multicolumn{1}{c}{\textbf{\scriptsize $\INC_C$}} &
\multicolumn{1}{c}{\textbf{\scriptsize $\INC_S$}} \\
\hline

\multirow{5}{*}{1-A} & HILL-0.40 & BOSS/BOSS & 500/500 & RM+EC &
0.2440 & 398 & 358 & 142 & 102 &
0.2410 & 0.1564 & 214 & 223 & 41 & 40 & 482 & 197 & 285 \\

& HILL-0.40 & BOSS/BOSS & 500/250 & RM+EC &
0.2573 & 199 & 358 & 142 & 51 &
0.2407 & 0.1491 & 108 & 223 & 41 & 17 & 361 & 169 & 192 \\

& HILL-0.40 & BOSS/BOSS & 500/0 & RM+EC &
0.2840 & 0 & 358 & 142 & 0 &
0.2360 & 0.1553 & 0 & 223 & 41 & 0 & 236 & 135 & 101 \\

& HILL-0.40 & BOSS/BOSS & 250/500 & RM+EC &
0.2320 & 398 & 178 & 72 & 102 &
0.2420 & 0.1628 & 214 & 110 & 23 & 40 & 363 & 130 & 233 \\

& HILL-0.40 & BOSS/BOSS & 0/500 & RM+EC &
0.2040 & 398 & 0 & 0 & 102 &
0.2460 & 0.1575 & 214 & 0 & 0 & 40 & 246 & 62 & 184 \\ \hdashline

\multirow{5}{*}{1-B} & HILL-0.40 & BOSS/BOWS2 & 500/500 & RM+EC &
0.4530 & 493 & 54 & 446 & 7 &
0.4365 & 0.7087 & 21 & 16 & 89 & 1 & 873 & 44 & 829 \\

& HILL-0.40 & BOSS/BOWS2 & 500/250 & RM+EC &
0.6027 & 244 & 54 & 446 & 6 &
0.4233 & 0.7826 & 9 & 16 & 89 & 1 & 635 & 43 & 592 \\

& HILL-0.40 & BOSS/BOWS2 & 500/0 & RM+EC &
0.8920 & 0 & 54 & 446 & 0 &
0.3950 & 0.8476 & 0 & 16 & 89 & 0 & 395 & 38 & 357 \\

& HILL-0.40 & BOSS/BOWS2 & 250/500 & RM+EC &
0.3067 & 493 & 27 & 223 & 7 &
0.4473 & 0.6203 & 21 & 9 & 48 & 1 & 671 & 24 & 647 \\

& HILL-0.40 & BOSS/BOWS2 & 0/500 & RM+EC &
0.0140 & 493 & 0 & 0 & 7 &
0.4780 & 0.0455 & 21 & 0 & 0 & 1 & 478 & 6 & 472 \\
\hdashline

\multirow{5}{*}{1-C} & HILL-0.40 & BOSS/ALASKA & 500/500 & RM+EC &
0.4810 & 369 & 150 & 350 & 131 &
0.4750 & 0.2600 & 19 & 18 & 11 & 2 & 950 & 261 & 689 \\

& HILL-0.40 & BOSS/ALASKA & 500/250 & RM+EC &
0.5453 & 191 & 150 & 350 & 59 &
0.4727 & 0.2927 & 11 & 18 & 11 & 1 & 709 & 190 & 519 \\

& HILL-0.40 & BOSS/ALASKA & 500/0 & RM+EC &
0.7000 & 0 & 150 & 350 & 0 &
0.4710 & 0.3793 & 0 & 18 & 11 & 0 & 471 & 132 & 339 \\

& HILL-0.40 & BOSS/ALASKA & 250/500 & RM+EC &
0.4027 & 369 & 79 & 171 & 131 &
0.4787 & 0.1875 & 19 & 7 & 4 & 2 & 718 & 201 & 517 \\

& HILL-0.40 & BOSS/ALASKA & 0/500 & RM+EC &
0.2620 & 369 & 0 & 0 & 131 &
0.4790 & 0.0952 & 19 & 0 & 0 & 2 & 479 & 129 & 350 \\
\hline

\multirow{5}{*}{2-A} & HILL-0.40 & BOSS/BOSS & 5000/5000 & RM+EC &
0.2515 & 3763 & 3722 & 1278 & 1237 &
0.2519 & 0.1477 & 2047 & 2183 & 341 & 392 & 5037 & 2384 & 2653 \\

& HILL-0.40 & BOSS/BOSS & 5000/2500 & RM+EC &
0.2511 & 1895 & 3722 & 1278 & 605 &
0.2499 & 0.1426 & 1034 & 2183 & 341 & 194 & 3748 & 1950 & 1798 \\

& HILL-0.40 & BOSS/BOSS & 5000/0 & RM+EC &
0.2556 & 0 & 3722 & 1278 & 0 &
0.2476 & 0.1351 & 0 & 2183 & 341 & 0 & 2476 & 1539 & 937 \\

& HILL-0.40 & BOSS/BOSS & 2500/5000 & RM+EC &
0.2483 & 3763 & 1875 & 625 & 1237 &
0.2541 & 0.1510 & 2047 & 1085 & 165 & 392 & 3811 & 1635 & 2176 \\

& HILL-0.40 & BOSS/BOSS & 0/5000 & RM+EC &
0.2474 & 3763 & 0 & 0 & 1237 &
0.2561 & 0.1607 & 2047 & 0 & 0 & 392 & 2561 & 845 & 1716 \\

\hdashline

\multirow{5}{*}{2-B} & HILL-0.40 & BOSS/BOWS2 & 5000/5000 & RM+EC &
0.4328 & 4814 & 858 & 4142 & 186 &
0.4426 & 0.6429 & 227 & 183 & 720 & 18 & 8852 & 843 & 8009 \\

& HILL-0.40 & BOSS/BOWS2 & 5000/2500 & RM+EC &
0.5640 & 2412 & 858 & 4142 & 88 &
0.4317 & 0.7090 & 115 & 183 & 720 & 6 & 6476 & 757 & 5719 \\

& HILL-0.40 & BOSS/BOWS2 & 5000/0 & RM+EC &
0.8284 & 0 & 858 & 4142 & 0 &
0.4097 & 0.7973 & 0 & 183 & 720 & 0 & 4097 & 675 & 3422 \\

& HILL-0.40 & BOSS/BOWS2 & 2500/5000 & RM+EC &
0.2980 & 4814 & 451 & 2049 & 186 &
0.4536 & 0.5460 & 227 & 89 & 362 & 18 & 6804 & 530 & 6274 \\

& HILL-0.40 & BOSS/BOWS2 & 0/5000 & RM+EC &
0.0372 & 4814 & 0 & 0 & 186 &
0.4755 & 0.0735 & 227 & 0 & 0 & 18 & 4755 & 168 & 4587 \\

\hdashline

\multirow{5}{*}{2-C} & HILL-0.40 & BOSS/ALASKA & 5000/5000 & RM+EC &
0.4826 & 3641 & 1533 & 3467 & 1359 &
0.4764 & 0.4386 & 146 & 119 & 147 & 60 & 9528 & 2713 & 6815 \\

& HILL-0.40 & BOSS/ALASKA & 5000/2500 & RM+EC &
0.5521 & 1826 & 1533 & 3467 & 674 &
0.4750 & 0.4693 & 80 & 119 & 147 & 29 & 7125 & 2059 & 5066 \\

& HILL-0.40 & BOSS/ALASKA & 5000/0 & RM+EC &
0.6934 & 0 & 1533 & 3467 & 0 &
0.4734 & 0.5526 & 0 & 119 & 147 & 0 & 4734 & 1414 & 3320 \\

& HILL-0.40 & BOSS/ALASKA & 2500/5000 & RM+EC &
0.4123 & 3641 & 767 & 1733 & 1359 &
0.4777 & 0.3970 & 146 & 56 & 73 & 60 & 7165 & 2010 & 5155 \\

& HILL-0.40 & BOSS/ALASKA & 0/5000 & RM+EC &
0.2718 & 3641 & 0 & 0 & 1359 &
0.4794 & 0.2913 & 146 & 0 & 0 & 60 & 4794 & 1299 & 3495 \\

\hline
\end{tabular}}
\end{center}
\end{table*}

\begin{table*}[ht]
\begin{center}

\caption{Classification results for standard RM+EC (without filters) and for the proposed method (with filters). Symbols and abbreviations have the same meaning as in Table \ref{tab:experiments}.}
\label{tab:experiments2}
\resizebox{0.975\textwidth}{!}{\begin{tabular}{lllcc|>{\bfseries}rrrrr|>{\bfseries}rrrrrrrrr}

\hline 

\multicolumn{5}{c|}{} &
\multicolumn{5}{c|}{\textbf{STANDARD}} &
\multicolumn{9}{c}{\textbf{PROPOSED}} \\

\textbf{\scriptsize N} &
\textbf{\scriptsize ALGO} & 
\textbf{\scriptsize DBs} &
\textbf{\scriptsize C/S} &
\textbf{\scriptsize CLF} &
\multicolumn{1}{c}{\textbf{\scriptsize Err}} & 
\multicolumn{1}{c}{\textbf{\scriptsize $\TP$}} & 
\multicolumn{1}{c}{\textbf{\scriptsize $\TN$}} & 
\multicolumn{1}{c}{\textbf{\scriptsize $\FP$}} & 
\multicolumn{1}{c|}{\textbf{\scriptsize $\FN$}} & 
\multicolumn{1}{c}{\textbf{\scriptsize Err$_{pred}$}} &
\multicolumn{1}{c}{\textbf{\scriptsize Err}} &
\multicolumn{1}{c}{\textbf{\scriptsize $\TP$}} & 
\multicolumn{1}{c}{\textbf{\scriptsize $\TN$}} & 
\multicolumn{1}{c}{\textbf{\scriptsize $\FP$}} & 
\multicolumn{1}{c}{\textbf{\scriptsize $\FN$}} &
\multicolumn{1}{c}{\textbf{\scriptsize $\INC$}} &
\multicolumn{1}{c}{\textbf{\scriptsize $\INC_C$}} &
\multicolumn{1}{c}{\textbf{\scriptsize $\INC_S$}} \\
\hline

\multirow{5}{*}{3-A} & UNIW-0.40 & BOSS/BOSS & 500/500 & RM+EC &
0.2030 & 419 & 378 & 122 & 81 &
0.1965 & 0.1203 & 257 & 277 & 40 & 33 & 393 & 149 & 244 \\

& UNIW-0.40 & BOSS/BOSS & 500/250 & RM+EC &
0.2240 & 204 & 378 & 122 & 46 &
0.1947 & 0.1201 & 126 & 277 & 40 & 15 & 292 & 132 & 160 \\

& UNIW-0.40 & BOSS/BOSS & 500/0 & RM+EC &
0.2440 & 0 & 378 & 122 & 0 &
0.1830 & 0.1262 & 0 & 277 & 40 & 0 & 183 & 101 & 82 \\

& UNIW-0.40 & BOSS/BOSS & 250/500 & RM+EC &
0.1920 & 419 & 187 & 63 & 81 &
0.1993 & 0.1175 & 257 & 141 & 20 & 33 & 299 & 94 & 205 \\

& UNIW-0.40 & BOSS/BOSS & 0/500 & RM+EC &
0.1620 & 419 & 0 & 0 & 81 &
0.2100 & 0.1138 & 257 & 0 & 0 & 33 & 210 & 48 & 162 \\
\hdashline

\multirow{5}{*}{3-B}& UNIW-0.40 & BOSS/BOW2 & 500/500 & RM+EC &
0.4330 & 485 & 82 & 418 & 15 &
0.3830 & 0.6795 & 43 & 32 & 157 & 2 & 766 & 63 & 703 \\

& UNIW-0.40 & BOSS/BOW2 & 500/250 & RM+EC &
0.5667 & 243 & 82 & 418 & 7 &
0.3573 & 0.7430 & 23 & 32 & 157 & 2 & 536 & 55 & 481 \\

& UNIW-0.40 & BOSS/BOW2 & 500/0 & RM+EC &
0.8360 & 0 & 82 & 418 & 0 &
0.3110 & 0.8307 & 0 & 32 & 157 & 0 & 311 & 50 & 261 \\

& UNIW-0.40 & BOSS/BOW2 & 250/500 & RM+EC &
0.2973 & 485 & 42 & 208 & 15 &
0.4080 & 0.5870 & 43 & 14 & 79 & 2 & 612 & 41 & 571 \\

& UNIW-0.40 & BOSS/BOW2 & 0/500 & RM+EC &
0.0300 & 485 & 0 & 0 & 15 &
0.4550 & 0.0444 & 43 & 0 & 0 & 2 & 455 & 13 & 442 \\
\hdashline

\multirow{5}{*}{3-C} & UNIW-0.40 & BOSS/ALASKA & 500/500 & RM+EC &
0.4830 & 397 & 120 & 380 & 103 &
0.4805 & 0.4103 & 12 & 11 & 16 & 0 & 961 & 212 & 749 \\

& UNIW-0.40 & BOSS/ALASKA & 500/250 & RM+EC &
0.5680 & 204 & 120 & 380 & 46 &
0.4780 & 0.4848 & 6 & 11 & 16 & 0 & 717 & 155 & 562 \\

& UNIW-0.40 & BOSS/ALASKA & 500/0 & RM+EC &
0.7600 & 0 & 120 & 380 & 0 &
0.4730 & 0.5926 & 0 & 11 & 16 & 0 & 473 & 109 & 364 \\

& UNIW-0.40 & BOSS/ALASKA & 250/500 & RM+EC &
0.3853 & 397 & 64 & 186 & 103 &
0.4867 & 0.2000 & 12 & 4 & 4 & 0 & 730 & 163 & 567 \\

& UNIW-0.40 & BOSS/ALASKA & 0/500 & RM+EC &
0.2060 & 397 & 0 & 0 & 103 &
0.4880 & 0.0000 & 12 & 0 & 0 & 0 & 488 & 103 & 385 \\

\hline

\multirow{5}{*}{4-A} & LSBm-0.10 & BOSS/BOSS & 500/500 & RM+EC &
0.0860 & 458 & 456 & 44 & 42 &
0.0780 & 0.0592 & 396 & 398 & 24 & 26 & 156 & 74 & 82 \\

& LSBm-0.10 & BOSS/BOSS & 500/250 & RM+EC &
0.0920 & 225 & 456 & 44 & 25 &
0.0767 & 0.0661 & 195 & 398 & 24 & 18 & 115 & 65 & 50 \\

& LSBm-0.10 & BOSS/BOSS & 500/0 & RM+EC &
0.0880 & 0 & 456 & 44 & 0 &
0.0780 & 0.0569 & 0 & 398 & 24 & 0 & 78 & 58 & 20 \\

& LSBm-0.10 & BOSS/BOSS & 250/500 & RM+EC &
0.0867 & 458 & 227 & 23 & 42 &
0.0760 & 0.0597 & 396 & 202 & 12 & 26 & 114 & 41 & 73 \\

& LSBm-0.10 & BOSS/BOSS & 0/500 & RM+EC &
0.0840 & 458 & 0 & 0 & 42 &
0.0780 & 0.0616 & 396 & 0 & 0 & 26 & 78 & 16 & 62 \\
\hdashline

\multirow{5}{*}{4-B} & LSBm-0.10 & BOSS/BOW2 & 500/500 & RM+EC &
0.1270 & 478 & 395 & 105 & 22 &
0.0975 & 0.0882 & 374 & 360 & 56 & 15 & 195 & 42 & 153 \\

& LSBm-0.10 & BOSS/BOW2 & 500/250 & RM+EC &
0.1520 & 241 & 395 & 105 & 9 &
0.0780 & 0.0979 & 211 & 360 & 56 & 6 & 117 & 38 & 79 \\

& LSBm-0.10 & BOSS/BOW2 & 500/0 & RM+EC &
0.2100 & 0 & 395 & 105 & 0 &
0.0840 & 0.1346 & 0 & 360 & 56 & 0 & 84 & 35 & 49 \\

& LSBm-0.10 & BOSS/BOW2 & 250/500 & RM+EC &
0.1227 & 478 & 180 & 70 & 22 &
0.1093 & 0.0939 & 374 & 157 & 40 & 15 & 164 & 30 & 134 \\

& LSBm-0.10 & BOSS/BOW2 & 0/500 & RM+EC &
0.0440 & 478 & 0 & 0 & 22 &
0.1110 & 0.0386 & 374 & 0 & 0 & 15 & 111 & 7 & 104 \\
\hdashline

\multirow{5}{*}{4-C} & LSBm-0.10 & BOSS/ALASKA & 500/500 & RM+EC &
0.4730 & 399 & 128 & 372 & 101 &
0.4500 & 0.5000 & 26 & 24 & 45 & 5 & 900 & 200 & 700 \\

& LSBm-0.10 & BOSS/ALASKA & 500/250 & RM+EC &
0.5667 & 197 & 128 & 372 & 53 &
0.4440 & 0.5595 & 13 & 24 & 45 & 2 & 666 & 155 & 511 \\

& LSBm-0.10 & BOSS/ALASKA & 500/0 & RM+EC &
0.7440 & 0 & 128 & 372 & 0 &
0.4310 & 0.6522 & 0 & 24 & 45 & 0 & 431 & 104 & 327 \\

& LSBm-0.10 & BOSS/ALASKA & 250/500 & RM+EC &
0.3867 & 399 & 61 & 189 & 101 &
0.4580 & 0.4127 & 26 & 11 & 21 & 5 & 687 & 146 & 541 \\

& LSBm-0.10 & BOSS/ALASKA & 0/500 & RM+EC &
0.2020 & 399 & 0 & 0 & 101 &
0.4690 & 0.1613 & 26 & 0 & 0 & 5 & 469 & 96 & 373 \\
\hline
\end{tabular}}
\end{center}
\end{table*}

Note that there is a case that cannot be detected with these filters: an image that is misclassified by all the classifiers. A graphical representation of this case is provided in Fig.\ref{fig:nasd_undetect}.

\section{Experimental results}
\label{sec:experimental}

The experimental validation of the method has been performed with different datasets selecting images from the following  databases:

\begin{itemize}

\item The BOSS database is the set of images from the Break Our Steganographic System! competition \cite{Bas:2011:boss}. This database is formed by 10,000 cover images taken with seven different cameras, with a size of $512\times512$ pixels. For JPEG experiments, we have compressed the images to qualities 75 and 95. In these cases, we refer to the datasets as BOSS-J75 and BOSS-J95, respectively.

\item The BOWS2 database is the set of images from the Break Our Watermarking System 2nd Ed. competition \cite{Bas:2007:bows2}. This database is formed by 10,000 cover images with a size of $512\times512$ pixels. For JPEG experiments, we have compressed the images to qualities 75 and 95. In these cases, we refer to the datasets as BOWS2-J75 and BOWS2-J95, respectively.

\item The ALASKA database is the set of images from the Alaska competition \cite{Cogranne:2018:alaska}. This database is formed by 50,000 cover images with different sizes. For our experiments, we have selected 10,000 images randomly and we  have cropped the center with a size of $512\times512$ pixels, and then converted them to gray scale. For JPEG experiments, we compressed the cropped images to qualities 75 and 95. In these cases, we refer to the datasets as ALASKA-J75 and ALASKA-J95, respectively.

\end{itemize}

For the experiments with SRNet \cite{Boroumand:2019:SRNet}, all these datasets have been resized to $256\times256$ pixels.

All the databases were randomly separated into two sets: a set for training and a set for testing. All the experiments, but Experiment 2 in Table \ref{tab:experiments}, use 9,500 images for training. Experiment 2 uses 5,000 images for training. The number of images used for testing varies among different experiments and it is provided in the different tables. These sets have been formed by the original cover images and the same images with an embedded message, using a random key. In this way, we have tested the CSM problem by using a training set from a database and a testing set from another one.

The experiments detailed below have been carried out for three different spatial domain steganographic algorithms: UNIversal \linebreak  WAvelet Relative Distortion (UNIWARD) \cite{Holub:2014:uniward}, HIgh-pass, Low-pass, and Low-pass (HILL) \cite{Li:2014:hill} and LSB matching (LSBM) \cite{Sharp:2001:lsbm}; and for two different transformed domain steganographic algorithms: UED \cite{Guo:2014:UED} and J-UNIWARD \cite{Holub:2014:uniward}. We were mainly interested in the results of the state-of-the-art algorithms, but we have also performed experiments using LSB matching to have a reference of a non-adaptive algorithm.

For these experiments, we have used a different random key for each stego image. The experiments have been performed using the same algorithm and embedding bit rate for the training and the testing sets unless explicitly noted otherwise. Thus, the proposed scenario is for a known algorithm and known embedding rate attack.
To analyze the proposed approach, we have used the well known  Rich Models (RM) framework \cite{Fridrich:2012:RM} for the spatial domain and Gabor Filter Residuals (GFR)  \cite{Song:2015:GFR} for the transformed domain, with Ensemble Classifiers (EC) \cite{Kodovsky:2012:EC}. These classifiers meet the requirements introduced in Section \ref{sec:proposed}, i.e., they make it possible to classify the sets $A\test$ (as $\CA$ and $\SA$) and $B\test$ (as $\SB$ and $\DB$) for UNIWARD, HILL, LSBM, UED and similar algorithms. When using RM+EC, we do not expect the proposed approach to work with algorithms designed to overcome this framework, such as \cite{Kouider:2013:aso}, because the proposed method depends on the underlying classifier.

We have also performed some experiments using the state-of-the-art convolutional neural network (CNN) described in \cite{Boroumand:2019:SRNet}.

\subsection{Classifier Inconsistencies}
\label{sec:inconsistent}

The results obtained detecting inconsistencies are presented in this section. In Tables 2-8, the results for different experiments are shown. The first column indicates the experiment number (used for references along the text). The first few columns show the algorithm and bit rate used for embedding, the databases used for training and for testing, the number of cover and stego images contained by the testing set, and the method used for classification. For example, the first row of the first experiment shows the results obtained using images from BOSS for training and images from the same database for testing (500 cover and 500 stego images), for HILL steganography with a 0.4 bpp embedding bit rate, using Rich Models with Ensemble Classifiers as the classification method. The results are computed first without applying any filter and next with the proposed filters. The column  error ($\Err$) is the result of the classification error using the reported  method. This error is computed as:
$$\Err=\frac{\FP+\FN}{\FP+\FN+\TP+\TN},$$
\noindent where $\TP$ are the true positives, $\TN$ the true negatives, $\FP$ the false positives and $\FN$ the false negatives. These values are also provided in Tables 2-8.

When the filters are applied, we also show
the number of inconsistencies ($\INC$) obtained during the classification, and which of those correspond to images classified by $\hat f_A$ as cover ($\INC_C$) or as stego ($\INC_S)$. When there is no CSM (for example, in the first row of Experiment 1-A of Table \ref{tab:experiments}) the number of inconsistencies is quite lower  compared to the cases with CSM (e.g. the first rows of Experiments 1-B  and 1-C of Table \ref{tab:experiments}).

It can also be observed that the number of inconsistencies increases with lower embedding bit rates (Experiment 5 of Table \ref{tab:experiments_lowerBR}). This occurs because the proposed method does not only reveal if there is CSM, but whether the classifier is working well with the testing database or not. This information can be used to make a good prediction about he classification errors, as detailed belown.

\subsection{Prediction of the Classification Error}
\label{sec:prederror}

In this section, we show how to carry out a prediction of the classification error. For this purpose, we assume that the standard steganalyzer is classifying randomly the samples that are not well represented by the classification model or by the training set. For example, in a balanced case with the same number of cover and stego images, a classifier that makes a classification error of $R$\%, is possibly having trouble with $2R$\% of the samples of testing set, but it is providing the correct class for half of the ``difficult'' samples by chance. The images belonging to these $2R$\% samples of the testing set are more likely to produce inconsistencies than those for which the classifier succeeds. For this reason, the number of inconsistencies tends to be similar to two times the number of errors produced by the standard classifier (without using filters). This fact can be observed in Table \ref{tab:experiments}. Note that, in the balanced case, $\INC_C$ is roughly two times the number of $\FN$ (obtained without filters) and $\INC_S$ is roughly two times the number of $\FP$ (also without filters).

This prediction makes it possible to approximate the classification error of the testing set with the standard classifier using the following expression:
$$\Err\pred = \frac{\INC}{2\left|A\test\right|},$$
\noindent where $\left|\cdot\right|$ denotes the cardinality of a set. A more general expression that can be applied in non-balanced cases is left for the future research. 

Tables 2-8 show the accuracy of the predicted classification error as compared to the true classification error. Please note that the predicted error can be computed without any knowledge of the true type (stego or cover) of the testing images, whereas the true classification error is computed using the true type of each image.
 
\begin{table*}[ht]
\begin{center}

\caption{Experiments with low bit rates. Symbols and abbreviations have the same meaning as in Table \ref{tab:experiments}.}
\label{tab:experiments_lowerBR}
\resizebox{0.975\textwidth}{!}{\begin{tabular}{lllcc|>{\bfseries}rrrrr|>{\bfseries}rrrrrrrrr}

\hline 

\multicolumn{5}{c|}{} &
\multicolumn{5}{c|}{\textbf{STANDARD}} &
\multicolumn{9}{c}{\textbf{PROPOSED}} \\

\textbf{\scriptsize N} &
\textbf{\scriptsize ALGO} & 
\textbf{\scriptsize DBs} &
\textbf{\scriptsize C/S} &
\textbf{\scriptsize CLF} &
\multicolumn{1}{c}{\textbf{\scriptsize Err}} & 
\multicolumn{1}{c}{\textbf{\scriptsize $\TP$}} & 
\multicolumn{1}{c}{\textbf{\scriptsize $\TN$}} & 
\multicolumn{1}{c}{\textbf{\scriptsize $\FP$}} & 
\multicolumn{1}{c|}{\textbf{\scriptsize $\FN$}} & 
\multicolumn{1}{c}{\textbf{\scriptsize Err$_{pred}$}} &
\multicolumn{1}{c}{\textbf{\scriptsize Err}} &
\multicolumn{1}{c}{\textbf{\scriptsize $\TP$}} & 
\multicolumn{1}{c}{\textbf{\scriptsize $\TN$}} & 
\multicolumn{1}{c}{\textbf{\scriptsize $\FP$}} & 
\multicolumn{1}{c}{\textbf{\scriptsize $\FN$}} &
\multicolumn{1}{c}{\textbf{\scriptsize $\INC$}} &
\multicolumn{1}{c}{\textbf{\scriptsize $\INC_C$}} &
\multicolumn{1}{c}{\textbf{\scriptsize $\INC_S$}} \\
\hline

\multirow{4}{*}{5} & HILL-0.20 & BOSS/BOSS & 500/500 & RM+EC &
0.3530 & 350 & 297 & 203 & 150 &
0.3545 & 0.2509 & 106 & 112 & 36 & 37 & 709 & 298 & 411 \\

& HILL-0.20 & BOSS/BOW2 & 500/500 & RM+EC &
0.4850 & 474 & 41 & 459 & 26 &
0.4875 & 0.7200 & 4 & 3 & 15 & 3 & 975 & 61 & 914 \\

& UNIW-0.20 & BOSS/BOSS & 500/500 & RM+EC &
0.3360 & 352 & 312 & 188 & 148 &
0.3205 & 0.1922 & 145 & 145 & 34 & 35 & 641 & 280 & 361 \\

& UNIW-0.20 & BOSS/BOW2 & 500/500 & RM+EC &
0.4650 & 485 & 50 & 450 & 15 &
0.4620 & 0.5658 & 19 & 14 & 40 & 3 & 924 & 48 & 876 \\
\hline

\hline
\end{tabular}}
\end{center}
\end{table*}

\begin{table*}[ht]
\begin{center}

\caption{Experiments training with images from other databases. Symbols and abbreviations have the same meaning as in Table \ref{tab:experiments}.}
\label{tab:experiments_otherDB}
\resizebox{0.975\textwidth}{!}{\begin{tabular}{lllcc|>{\bfseries}rrrrr|>{\bfseries}rrrrrrrrr}

\hline 

\multicolumn{5}{c|}{} &
\multicolumn{5}{c|}{\textbf{STANDARD}} &
\multicolumn{9}{c}{\textbf{PROPOSED}} \\

\textbf{\scriptsize N} &
\textbf{\scriptsize ALGO} & 
\textbf{\scriptsize DBs} &
\textbf{\scriptsize C/S} &
\textbf{\scriptsize CLF} &
\multicolumn{1}{c}{\textbf{\scriptsize Err}} & 
\multicolumn{1}{c}{\textbf{\scriptsize $\TP$}} & 
\multicolumn{1}{c}{\textbf{\scriptsize $\TN$}} & 
\multicolumn{1}{c}{\textbf{\scriptsize $\FP$}} & 
\multicolumn{1}{c|}{\textbf{\scriptsize $\FN$}} & 
\multicolumn{1}{c}{\textbf{\scriptsize Err$_{pred}$}} &
\multicolumn{1}{c}{\textbf{\scriptsize Err}} &
\multicolumn{1}{c}{\textbf{\scriptsize $\TP$}} & 
\multicolumn{1}{c}{\textbf{\scriptsize $\TN$}} & 
\multicolumn{1}{c}{\textbf{\scriptsize $\FP$}} & 
\multicolumn{1}{c}{\textbf{\scriptsize $\FN$}} &
\multicolumn{1}{c}{\textbf{\scriptsize $\INC$}} &
\multicolumn{1}{c}{\textbf{\scriptsize $\INC_C$}} &
\multicolumn{1}{c}{\textbf{\scriptsize $\INC_S$}} \\
\hline


\multirow{3}{*}{6-A} & HILL-0.40 & BOWS2/BOWS2 & 500/500 & RM+EC &
0.2060 & 414 & 380 & 120 & 86 &
0.1900 & 0.1161 & 268 & 280 & 34 & 38 & 380 & 148 & 232 \\

& HILL-0.40 & BOWS2/BOSS & 500/500 & RM+EC &
0.3140 & 364 & 322 & 178 & 136 &
0.3055 & 0.2031 & 148 & 162 & 42 & 37 & 611 & 259 & 352 \\

& HILL-0.40 & BOWS2/ALASKA & 500/500 & RM+EC &
0.4710 & 363 & 166 & 334 & 137 &
0.4685 & 0.3968 & 20 & 18 & 17 & 8 & 937 & 277 & 660 \\

\hdashline

\multirow{3}{*}{6-B}  & HILL-0.40 & ALASKA/ALASKA & 500/500 & RM+EC &
0.2950 & 378 & 327 & 173 & 122 &
0.2895 & 0.1354 & 177 & 187 & 31 & 26 & 579 & 236 & 343 \\

& HILL-0.40 & ALASKA/BOSS & 500/500 & RM+EC &
0.4110 & 387 & 202 & 298 & 113 &
0.3905 & 0.2466 & 78 & 87 & 27 & 27 & 781 & 201 & 580 \\

& HILL-0.40 & ALASKA/BOWS2 & 500/500 & RM+EC &
0.3710 & 273 & 356 & 144 & 227 &
0.3730 & 0.2756 & 81 & 103 & 29 & 41 & 746 & 439 & 307 \\

\hline
\end{tabular}}
\end{center}
\end{table*}

\begin{table*}[ht]
\begin{center}

\caption{Experiments using the SRNet convolutional neural network. Symbols and abbreviations have the same meaning as in Table \ref{tab:experiments}.}
\label{tab:experiments_srnet}
\resizebox{0.975\textwidth}{!}{\begin{tabular}{lllcc|>{\bfseries}rrrrr|>{\bfseries}rrrrrrrrr}

\hline 

\multicolumn{5}{c|}{} &
\multicolumn{5}{c|}{\textbf{STANDARD}} &
\multicolumn{9}{c}{\textbf{PROPOSED}} \\

\textbf{\scriptsize N} &
\textbf{\scriptsize ALGO} & 
\textbf{\scriptsize DBs} &
\textbf{\scriptsize C/S} &
\textbf{\scriptsize CLF} &
\multicolumn{1}{c}{\textbf{\scriptsize Err}} & 
\multicolumn{1}{c}{\textbf{\scriptsize $\TP$}} & 
\multicolumn{1}{c}{\textbf{\scriptsize $\TN$}} & 
\multicolumn{1}{c}{\textbf{\scriptsize $\FP$}} & 
\multicolumn{1}{c|}{\textbf{\scriptsize $\FN$}} & 
\multicolumn{1}{c}{\textbf{\scriptsize Err$_{pred}$}} &
\multicolumn{1}{c}{\textbf{\scriptsize Err}} &
\multicolumn{1}{c}{\textbf{\scriptsize $\TP$}} & 
\multicolumn{1}{c}{\textbf{\scriptsize $\TN$}} & 
\multicolumn{1}{c}{\textbf{\scriptsize $\FP$}} & 
\multicolumn{1}{c}{\textbf{\scriptsize $\FN$}} &
\multicolumn{1}{c}{\textbf{\scriptsize $\INC$}} &
\multicolumn{1}{c}{\textbf{\scriptsize $\INC_C$}} &
\multicolumn{1}{c}{\textbf{\scriptsize $\INC_S$}} \\
\hline

\multirow{3}{*}{7-A}  & HILL-0.40 & BOSS/BOSS & 500/500 & SRNET &
0.2520 & 434 & 314 & 186 & 66 &
0.2635 & 0.1057 & 216 & 207 & 35 & 15 & 527 & 158 & 369 \\

& HILL-0.40 & BOSS/BOWS2 & 500/500 & SRNET &
0.2600 & 447 & 293 & 207 & 53 &
0.2855 & 0.1235 & 178 & 198 & 33 & 20 & 571 & 128 & 443 \\

& HILL-0.40 & BOSS/ALASKA & 500/500 & SRNET &
0.3840 & 441 & 175 & 325 & 59 &
0.3825 & 0.2128 & 75 & 110 & 22 & 28 & 765 & 96 & 669 \\

\hdashline

\multirow{3}{*}{7-B} & HILL-0.40 & BOWS2/BOWS2 & 500/500 & SRNET &
0.2670 & 437 & 296 & 204 & 63 &
0.2720 & 0.1272 & 184 & 214 & 35 & 23 & 544 & 122 & 422 \\

& HILL-0.40 & BOWS2/BOSS & 500/500 & SRNET &
0.3570 & 452 & 191 & 309 & 48 &
0.3250 & 0.2000 & 156 & 124 & 62 & 8 & 650 & 107 & 543 \\

& HILL-0.40 & BOWS2/ALASKA & 500/500 & SRNET &
0.3880 & 319 & 293 & 207 & 181 &
0.3805 & 0.2218 & 80 & 106 & 28 & 25 & 761 & 343 & 418 \\

\hdashline

\multirow{3}{*}{7-C} & HILL-0.40 & ALASKA/ALASKA & 500/500 & SRNET &
0.3940 & 324 & 282 & 218 & 176 &
0.3910 & 0.2156 & 76 & 95 & 34 & 13 & 782 & 350 & 432 \\

& HILL-0.40 & ALASKA/BOWS2 & 500/500 & SRNET &
0.3930 & 438 & 169 & 331 & 62 &
0.3930 & 0.2150 & 77 & 91 & 35 & 11 & 786 & 129 & 657 \\

& HILL-0.40 & ALASKA/BOSS & 500/500 & SRNET &
0.3900 & 386 & 224 & 276 & 114 &
0.4120 & 0.2955 & 68 & 56 & 43 & 9 & 824 & 273 & 551 \\

\hline
\end{tabular}}
\end{center}
\end{table*}

\begin{table*}[ht]
\begin{center}

\caption{Experiments with unknown bitrate (SSM). Symbols and abbreviations have the same meaning as in Table \ref{tab:experiments}.}
\label{tab:experiments_ssm}
\resizebox{0.975\textwidth}{!}{\begin{tabular}{lllcc|>{\bfseries}rrrrr|>{\bfseries}rrrrrrrrr}

\hline 

\multicolumn{5}{c|}{} &
\multicolumn{5}{c|}{\textbf{STANDARD}} &
\multicolumn{9}{c}{\textbf{PROPOSED}} \\

\textbf{\scriptsize N} &
\textbf{\scriptsize ALGO} & 
\textbf{\scriptsize DBs} &
\textbf{\scriptsize C/S} &
\textbf{\scriptsize CLF} &
\multicolumn{1}{c}{\textbf{\scriptsize Err}} & 
\multicolumn{1}{c}{\textbf{\scriptsize $\TP$}} & 
\multicolumn{1}{c}{\textbf{\scriptsize $\TN$}} & 
\multicolumn{1}{c}{\textbf{\scriptsize $\FP$}} & 
\multicolumn{1}{c|}{\textbf{\scriptsize $\FN$}} & 
\multicolumn{1}{c}{\textbf{\scriptsize Err$_{pred}$}} &
\multicolumn{1}{c}{\textbf{\scriptsize Err}} &
\multicolumn{1}{c}{\textbf{\scriptsize $\TP$}} & 
\multicolumn{1}{c}{\textbf{\scriptsize $\TN$}} & 
\multicolumn{1}{c}{\textbf{\scriptsize $\FP$}} & 
\multicolumn{1}{c}{\textbf{\scriptsize $\FN$}} &
\multicolumn{1}{c}{\textbf{\scriptsize $\INC$}} &
\multicolumn{1}{c}{\textbf{\scriptsize $\INC_C$}} &
\multicolumn{1}{c}{\textbf{\scriptsize $\INC_S$}} \\
\hline


\multirow{5}{*}{8} & HILL-0.40/0.20 & BOSS/BOSS & 500/500 & RM+EC &
0.2850 & 418 & 297 & 203 & 82 &
0.4200 & 0.2625 & 6 & 112 & 36 & 6 & 840 & 261 & 579 \\

& HILL-0.40/0.30 & BOSS/BOSS & 500/500 & RM+EC &
0.2450 & 413 & 342 & 158 & 87 &
0.3470 & 0.1667 & 84 & 171 & 35 & 16 & 694 & 242 & 452 \\

& HILL-0.40/0.40 & BOSS/BOSS & 500/500 & RM+EC &
0.2440 & 398 & 358 & 142 & 102 &
0.2410 & 0.1564 & 214 & 223 & 41 & 40 & 482 & 197 & 285 \\

& HILL-0.40/0.50 & BOSS/BOSS & 500/500 & RM+EC &
0.2640 & 344 & 392 & 108 & 156 &
0.1880 & 0.1538 & 250 & 278 & 35 & 61 & 376 & 209 & 167 \\

& HILL-0.40/0.60 & BOSS/BOSS & 500/500 & RM+EC &
0.2820 & 312 & 406 & 94 & 188 &
0.1800 & 0.1422 & 242 & 307 & 33 & 58 & 360 & 229 & 131 \\

\hline
\end{tabular}}
\end{center}
\end{table*}

We have carried out experiments with different ratios of cover and stego images (Experiments 1-2 in Table \ref{tab:experiments}, and 3-4 in Table \ref{tab:experiments2}). In the balanced case (500/500), the predicted error ($\Err\pred$) is very close to the true classification error ($\Err$). However, in the case of unbalanced number of cover and stego images, the predicted classification error is not that close to the real value, but it is close to the true classification error obtained with a balanced testing set.

Note that, due to the proposed formula, the prediction of the classification error will always be between 0 and $0.5$. Therefore, a classifier that does not work for a given testing set would yield a prediction of the classification error about $0.5$. In unbalanced experiments, we can obtain true classification errors above $0.5$, such as in the third row of the Experiment 1-C (Table \ref{tab:experiments}), with an error of $0.7000$. Our prediction is $0.4787$ indicating that the classifier is random guessing with this testing set. A similar situation occurs with apparently good classification results, as in the last row of Experiment 1-B (Table \ref{tab:experiments}). The classification error is $0.0140$, whereas the prediction of the classification error is $0.4790$. Note that the predicted classification error is correct, as far as the classifier is not working in these conditions, as evident from the balanced case (first row of Experiment 1-B). The reason why the classification error is so small in the all-stego case is that the output of the classifier is stego for almost all images, which works by chance when all testing images are stego.

Note that Experiment 2 (Table \ref{tab:experiments}) is the same as Experiment 1 (Table \ref{tab:experiments}) but using 10,000 images for training and 10,000 images for testing. This experiment was carried out to check if the results with 1,000 images are stable enough. We can see that, in both cases, the results are very similar.

In Experiments 3 and 4 (Table \ref{tab:experiments2}), the results obtained for the embedding algorithms UNIWARD and LSB matching are shown, and the accuracy of the predicted classification error is similar to that of Table \ref{tab:experiments}. Comparable results are also obtained with low embedding bit rates, as it can be observed in Experiment 5 (Table \ref{tab:experiments_lowerBR}) for the algorithms HILL and UNIWARD with $0.2$ bpp.

In the previous experiments, the database used for training is BOSS. In Experiment 6 (Table \ref{tab:experiments_otherDB}), we show the results obtained using images from other databases for training. More precisely, we use BOWS2 and ALASKA in the training set. As it can be observed, the results are comparable in terms of the accuracy of the prediction of the classification error. 

In Experiment 7 (Table \ref{tab:experiments_srnet}), the results using the SRNet \cite{Boroumand:2019:SRNet}  classification method are presented, and different training databases are used. Again, the prediction of the classification errors is accurate.

Experiment 8 (Table \ref{tab:experiments_ssm}) presents the classification results in case of SSM. In this case, the testing set is embedded with HILL and 0.40 bpb, and the  embedding bit rate of the training set varies between 0.20 and 0.60 bpp. When a wrong embedding bit rate is chosen, the prediction of the classification error is less accurate. The reason for this mismatch is that a wrong embedding rate is used to create the set $B\train$ and, hence, the classifier ${\hat f}_B$ is not appropriate for the testing set. This problem will be addressed in our future work.

Finally, in Experiments 9 and 10 (Table \ref{tab:experiments_JPEG}), we show the results obtained for JPEG images compressed to qualities 75 and 95, respectively. In this case, we have used the embedding algorithms UED and J-UNIWARD, and the accuracy of the predicted classification error is consistent with that of the rest of the experiments.

\begin{table*}[ht]
\begin{center}

\caption{Experiments with JPEG steganography. Symbols and abbreviations have the same meaning as in Table \ref{tab:experiments}.}
\label{tab:experiments_JPEG}
\resizebox{0.975\textwidth}{!}{\begin{tabular}{lllcc|>{\bfseries}rrrrr|>{\bfseries}rrrrrrrrr}

\hline 

\multicolumn{5}{c|}{} &
\multicolumn{5}{c|}{\textbf{STANDARD}} &
\multicolumn{9}{c}{\textbf{PROPOSED}} \\

\textbf{\scriptsize N} &
\textbf{\scriptsize ALGO} & 
\textbf{\scriptsize DBs} &
\textbf{\scriptsize C/S} &
\textbf{\scriptsize CLF} &
\multicolumn{1}{c}{\textbf{\scriptsize Err}} & 
\multicolumn{1}{c}{\textbf{\scriptsize $\TP$}} & 
\multicolumn{1}{c}{\textbf{\scriptsize $\TN$}} & 
\multicolumn{1}{c}{\textbf{\scriptsize $\FP$}} & 
\multicolumn{1}{c|}{\textbf{\scriptsize $\FN$}} & 
\multicolumn{1}{c}{\textbf{\scriptsize Err$_{pred}$}} &
\multicolumn{1}{c}{\textbf{\scriptsize Err}} &
\multicolumn{1}{c}{\textbf{\scriptsize $\TP$}} & 
\multicolumn{1}{c}{\textbf{\scriptsize $\TN$}} & 
\multicolumn{1}{c}{\textbf{\scriptsize $\FP$}} & 
\multicolumn{1}{c}{\textbf{\scriptsize $\FN$}} &
\multicolumn{1}{c}{\textbf{\scriptsize $\INC$}} &
\multicolumn{1}{c}{\textbf{\scriptsize $\INC_C$}} &
\multicolumn{1}{c}{\textbf{\scriptsize $\INC_S$}} \\
\hline


\multirow{5}{*}{9-A} & UED-0.40 & BOSS-J75/BOSS-J75 & 500/500 & GFR+EC &
0.0290 & 480 & 491 & 9 & 20 &
0.0315 & 0.0267 & 452 & 460 & 8 & 17 & 63 & 34 & 29 \\

& UED-0.40 & BOSS-J75/BOWS2-J75 & 500/500 & GFR+EC &
0.0300 & 481 & 489 & 11 & 19 &
0.0355 & 0.0215 & 450 & 459 & 6 & 14 & 71 & 35 & 36 \\

& UED-0.40 & BOSS-J75/ALASKA-J75 & 500/500 & GFR+EC &
0.2090 & 348 & 443 & 57 & 152 &
0.2315 & 0.1899 & 204 & 231 & 29 & 73 & 463 & 291 & 172 \\

& J-UNIW-0.40 & BOSS-J75/BOSS-J75 & 500/500 & GFR+EC &
0.0820 & 446 & 472 & 28 & 54 &
0.0910 & 0.0819 & 362 & 389 & 22 & 45 & 182 & 92 & 90 \\

& J-UNIW-0.40 & BOSS-J75/BOWS2-J75 & 500/500 & GFR+EC &
0.1000 & 445 & 455 & 45 & 55 &
0.0990 & 0.0998 & 348 & 374 & 37 & 43 & 198 & 93 & 105 \\

\hdashline
\multirow{5}{*}{9-B} & UED-0.40 & BOWS2-J75/BOWS2-J75 & 500/500 & GFR+EC &
0.0240 & 490 & 486 & 14 & 10 &
0.0340 & 0.0193 & 452 & 462 & 9 & 9 & 68 & 25 & 43 \\

& UED-0.40 & BOWS2-J75/BOSS-J75 & 500/500 & GFR+EC &
0.0350 & 487 & 478 & 22 & 13 &
0.0305 & 0.0309 & 452 & 458 & 19 & 10 & 61 & 23 & 38 \\

& UED-0.40 & BOWS2-J75/ALASKA-J75 & 500/500 & GFR+EC &
0.2070 & 346 & 447 & 53 & 154 &
0.2435 & 0.1910 & 194 & 221 & 27 & 71 & 487 & 309 & 178 \\

& J-UNIW-0.40 & BOWS2-J75/BOWS2-J75 & 500/500 & GFR+EC &
0.0970 & 450 & 453 & 47 & 50 &
0.0990 & 0.0960 & 352 & 373 & 35 & 42 & 198 & 88 & 110 \\

& J-UNIW-0.40 & BOWS2-J75/BOSS-J75 & 500/500 & GFR+EC &
0.0960 & 466 & 438 & 62 & 34 &
0.1165 & 0.0769 & 346 & 362 & 38 & 21 & 233 & 89 & 144 \\

\hdashline
\multirow{3}{*}{9-C} & UED-0.40 & ALASKA-J75/ALASKA-J75 & 500/500 & GFR+EC &
0.0800 & 456 & 464 & 36 & 44 &
0.0955 & 0.0581 & 375 & 387 & 20 & 27 & 191 & 94 & 97 \\

& UED-0.40 & ALASKA-J75/BOSS-J75 & 500/500 & GFR+EC &
0.0680 & 453 & 479 & 21 & 47 &
0.0805 & 0.0536 & 385 & 409 & 16 & 29 & 161 & 88 & 73 \\

& UED-0.40 & ALASKA-J75/BOWS2-J75 & 500/500 & GFR+EC &
0.0890 & 442 & 469 & 31 & 58 &
0.0925 & 0.0847 & 363 & 383 & 24 & 45 & 185 & 99 & 86 \\

\hline

\multirow{5}{*}{10-A} & UED-0.40 & BOSS-J95/BOSS-J95 & 500/500 & GFR+EC &
0.1530 & 430 & 417 & 83 & 70 &
0.1310 & 0.1220 & 319 & 329 & 47 & 43 & 262 & 115 & 147 \\

& UED-0.40 & BOSS-J95/BOWS2-J95 & 500/500 & GFR+EC &
0.1900 & 368 & 442 & 58 & 132 &
0.1635 & 0.1842 & 267 & 282 & 30 & 94 & 327 & 198 & 129 \\

& UED-0.40 & BOSS-J95/ALASKA-J95 & 500/500 & GFR+EC &
0.4310 & 140 & 429 & 71 & 360 &
0.4145 & 0.4152 & 44 & 56 & 13 & 58 & 829 & 675 & 154 \\

& J-UNIW-0.40 & BOSS-J95/BOSS-J95 & 500/500 & GFR+EC &
0.2280 & 369 & 403 & 97 & 131 &
0.2295 & 0.2089 & 201 & 227 & 50 & 63 & 459 & 244 & 215 \\

& J-UNIW-0.40 & BOSS-J95/BOWS2-J95 & 500/500 & GFR+EC &
0.2640 & 324 & 412 & 88 & 176 &
0.2560 & 0.2295 & 183 & 193 & 37 & 75 & 512 & 320 & 192 \\

\hdashline

\multirow{5}{*}{10-B} & UED-0.40 & BOWS2-J95/BOWS2-J95 & 500/500 & GFR+EC &
0.1660 & 414 & 420 & 80 & 86 &
0.1525 & 0.1640 & 285 & 296 & 47 & 67 & 305 & 143 & 162 \\

& UED-0.40 & BOWS2-J95/BOSS-J95 & 500/500 & GFR+EC &
0.1690 & 427 & 404 & 96 & 73 &
0.1460 & 0.1427 & 306 & 301 & 52 & 49 & 292 & 127 & 165 \\

& UED-0.40 & BOWS2-J95/ALASKA-J95 & 500/500 & GFR+EC &
0.4180 & 154 & 428 & 72 & 346 &
0.3995 & 0.3980 & 53 & 68 & 22 & 58 & 799 & 648 & 151 \\

& J-UNIW-0.40 & BOWS2-J95/BOWS2-J95 & 500/500 & GFR+EC &
0.2600 & 366 & 374 & 126 & 134 &
0.2380 & 0.2481 & 194 & 200 & 71 & 59 & 476 & 249 & 227 \\

& J-UNIW-0.40 & BOWS2-J95/BOSS-J95 & 500/500 & GFR+EC &
0.2460 & 380 & 374 & 126 & 120 &
0.2380 & 0.2252 & 194 & 212 & 59 & 59 & 476 & 223 & 253 \\

\hdashline
\multirow{3}{*}{10-C} & UED-0.40 & ALASKA-J95/ALASKA-J95 & 500/500 & GFR+EC &
0.3040 & 349 & 347 & 153 & 151 &
0.2665 & 0.2227 & 183 & 180 & 35 & 69 & 533 & 249 & 284 \\

& UED-0.40 & ALASKA-J95/BOSS-J95 & 500/500 & GFR+EC &
0.2350 & 390 & 375 & 125 & 110 &
0.2065 & 0.2095 & 229 & 235 & 59 & 64 & 413 & 186 & 227 \\

& UED-0.40 & ALASKA-J95/BOWS2-J95 & 500/500 & GFR+EC &
0.2400 & 356 & 404 & 96 & 144 &
0.2100 & 0.2224 & 215 & 236 & 41 & 88 & 420 & 224 & 196 \\
\hline

\end{tabular}}
\end{center}
\end{table*}


As shown in the experiments, the proposed method works both if there is CSM and when it is too difficult to classify images with the underlying classifier in case of a too small embedding bit rate.

\section{Conclusion}
\label{sec:conclusion}

In this paper, a method for detecting inconsistencies in image {steg\-analysis} is presented.
We show how the number of inconsistencies can be used to predict the classification error of the steganalytic method. The proposed approach has been tested for different steganalyzers, image databases, embedding algorithms and embedding bit rates, with and without CSM.


The results show how the classification error of a steganalytic method can be predicted without having access to the labels of the images in the testing set. The predicted classification error can be a very valuable information for a steganalyst, who can decide how to proceed when the predicted classification error is too large. In such a case, increasing the training set in order to improve the classification accuracy could be one of the alternatives to be considered.

The proposed method is intended to be used in batch steganography. Nevertheless, even for a single testing image, this approach makes it possible to detect if the classification is inconsistent. In such a case, the classifier should not be used to classify that image. As future work, it would be worth analyzing how to take profit of the proposed methodology when classifying single images. Finally, in case of \emph{stego source mismatch} (e.g. when the embedding bit rate is not known accurately) the prediction of the classification is not reliable. Possible approaches to deal with this problem will be addressed in our future research.

\begin{acks}
This work was supported by the Spanish Government, in part under Grant RTI2018-095094-B-C22 ``CONSENT'', and in part under Grant TIN2014-57364-C2-2-R ``SMARTGLACIS.''

We gratefully acknowledge the support of NVIDIA Corporation with the donation of an NVIDIA TITAN Xp GPU card that has been used in this work.
\end{acks}

\bibliographystyle{ACM-Reference-Format}
\bibliography{main}



\end{document}